\documentstyle[aps,multicol,epsfig,prl,fancyheadings]{revtex}

\begin{document}
\bibliographystyle{prsty}
\input epsf
\draft \preprint{HEP/123-qed}
\title{Ratcheting of granular materials}
\author{F. Alonso-Marroqu\'{\i}n, H. J. Herrmann}
\address{ICA1, University of Stuttgart, \\
 Pfaffenwaldring 27, 70569 Stuttgart, Germany\\
}

\maketitle

\begin{abstract}

We investigate the quasi-static mechanical response of soils under cyclic 
loading using a discrete model of randomly generated convex polygons. This
response exhibits a sequence of regimes, each one  characterized by a 
linear accumulation of plastic deformation with the number of cycles. 
At the grain level, a quasi-periodic ratchet-like  
behavior is observed at the  contacts, which  excludes the 
existence of an elastic regime. The study of this slow dynamics allows 
to explore the role of  friction in the permanent deformation of unbound 
granular materials supporting railroads and  streets.


\end{abstract}

\begin{multicols}{2}

A particularly intriguing phenomenon in driven systems is the so-called
ratchet effect. There is already an extensive body of work on this
subject, driven by the need to understand biophysical systems such as 
molecular motors \cite{howard97} and  certain mechanical 
and electrical rectifiers \cite{reimann02}. 
The classic ratchet is a mechanical device consisting of  a pawl that
engages the sloping  teeth of a  wheel, permitting  motion in one
direction only. Ratchet-like motion have been proposed as a mechanism
to explain the convective motion and size segregation in vibrated 
granular materials \cite{jaeger96}. The understanding of this phenomenon 
is crucial in the investigation on the permanent deformation in structures 
subjected to  cyclic loading. In soils, this loading can be induced by 
earthquakes,  sea waves, road traffic, etc. \cite{lekarp00}.

The classical theory of elasto-plasticity describes the cyclic loading
response by postulating an elastic region in the stress space,  which 
changes during the deformation \cite{drucker52}. 
This elastic region,  however, is not easy to 
identify because the onset of the plastic deformation is gradual and
not sharply defined. A great variety of modifications have been proposed in
order to provide a more appropriate description 
which, however, make the theory too complicated, and  require
too many material parameters which are difficult to calibrate
\cite{gudehus84}. 

This research was motivated by  experiments  of cyclic  
loading tests on unbound granular material used to support 
railroads and streets. A slow deformation is observed 
during the service life of these structures, where the breakage, corrosion 
and  the friction between the grains play an important role
\cite{lekarp00}. We emphasise some recent experiments 
performed in Darmstadt  \cite{festag02}. These experiments show that 
when the samples  consist of very wear resistant grains, the long time 
cyclic loading behavior is given by a linear accumulation of plastic  
deformation.
This surprising result suggests that the grains attain a periodic 
irreversible motion at the sliding contacts, which could, in principle,
be detected  using numerical simulations.  
Here we report on the first micromechanical  observation 
using molecular dynamic simulations.  Ratcheting motion was detected in 
the sliding  contacts on a polygonal packing subjected to  quasi-static 
cyclic loading.  

The polygons representing  the particles in this model are generated
by using a  simple version of Voronoi tessellation: 
First, we set a random point in each cell of a regular square 
lattice of side $\ell$, then each polygon is constructed assigning to 
each point  that part of  the plane that is nearer to it than  
to any other point.  This method gives a diversity of areas of 
polygons  following a Gaussian distribution with mean value $\ell^2$ 
and variance  of $0.36\ell^2$. The number of edges of the polygons is 
distributed between  $4$ and $8$ for $98.7\%$ of the polygons, 
with a mean value of $6$.

The interaction between the polygons is modelled as follows: 
when two polygons overlap,  two points can be defined  by the intersection 
of their edges. The segment connecting these two intersection points 
defines the contact line. 
Then, the elastic part of the contact force is  calculated as 
$\vec{f^e}=k_n \Delta x^e_n \hat{n}^c + k_t \Delta x^e_t \hat{t}^c$,
where $\hat{n}^c$ and $\hat{t}^c$ denote the normal and tangential unitary
vectors with respect to the contact line, and $k_n$ and $k_t$ are the
stiffnesses in the respective directions.  
$\Delta x^e_n$  is the overlapping area 
divided by a characteristic  length of the interacting polygon pair 
\cite{kun99}. 
$\Delta x^e_t$ defines the elastic tangential displacement of the contact, 
that is given by the time integral starting at the beginning of the contact

\begin{equation}
\Delta x^e_t=\int_{0}^{t}v^c_t(t')\Theta(\mu f^e_n-|f^e_t|)dt',
\label{friction} 
\end{equation}

\noindent
where $\Theta$ is the Heaviside step function and $\vec{v}^c_t$ denotes the 
tangential component of the relative velocity $\vec{v}^c$ at the contact
\cite{alonso02a}.
This equation defines a limit of elasticity in the contact force. When 
the contact force satisfies $f^e_t = \mu f^e_n$  the contact slides.
This produces an irreversible motion between the polygons which contributes 
to a plastic deformation in the assembly. 

In each contact a viscous force $\vec{f}^v = m \nu \vec{v}^c$ is included. 
Here $m$ is  the relative mass of the polygons in contact and $\nu$ is the 
coefficient of viscosity.  This force reduces the acoustic waves produced 
during the loading.   Inertia effects were excluded by using a very long 
time of loading. indeed, doubling it affects the strain response by less 
than $5\%$. This correspond to the quasi-static approximation.

\begin{figure}[h]
\begin{center}
  \epsfig{file=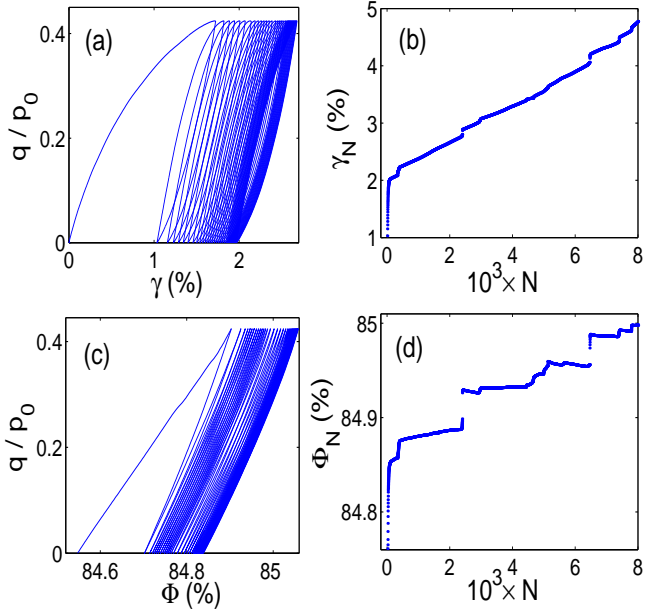,height=0.95\linewidth,width=1.0\linewidth,angle=0,clip=1}
 \end{center}
\caption{(a) Shear stress versus shear strain  in the first $40$ cycles. (b) 
permanent (plastic) strain $\gamma_N$ after $N$ cycles versus the number 
of cycles. 
(c) stress against the volume fraction in the first $40$ cycles. 
(d) volume fraction $\Phi_N$ after $N$ cycles versus number of cycles.}
\label{mosaico}
\end{figure}

The parameters of our model have been reduced to a minimum set of
dimensionless parameters:  The period of cyclic loading $t_0$ is chosen
$4000$ times the characteristic period of oscillation 
$t_s=\sqrt{k_n/\rho\ell^2}$, where $k_n$ is the normal stiffness of 
the contacts and $\rho$  the density of the grains. The  relaxation time 
$t_r=1/\nu$ is chosen 
$10$ times $t_s$.   
The tangential stiffness is  $k_t = 0.33 k_n$;  the initial pressure 
$p_0=0.001 k_n$ and the friction  coefficient $\mu =0.25$. The normal
stiffness is chossen $k_n = 160 MPa$.

The simulations are performed on six different samples of $400$ 
polygons.   First, the polygons are placed into a rectangular box
such that they do not overlap each other. Next, a gravitational
field is applied and the sample is allowed to consolidate.  
An external load is imposed by  applying a 
force $\sigma_1 W$ and $\sigma_2 H$ on the  horizontal and 
vertical walls, respectively. Here  $\sigma_1$ and $\sigma_2$ are the
horizontal and vertical stresses. $W$  and $H$ are the width and the height 
of the sample. When the velocity of the polygons vanishes gravity is 
switched off. A fifth-order  predictor-corrector algorithm is used to 
solve the equations of motion.

In order to calculate the strain we select the polygons whose centers of 
mass are less than $10 \ell$ from the center of the sample. 
Then, the strain tensor is  calculated  as the displacement gradient tensor
averaged over the area enclosed by the initial configuration of these 
polygons. From the eigenvalues  $\epsilon_1$ and $\epsilon_2$ of the 
symmetric part of this tensor we obtain the shear strain as 
$\gamma=\epsilon_1-\epsilon_2$. 
The volume  fraction  is calculated as $\Phi=(V_p-V_0)/V_b$, where $V_p$ 
is the sum of the areas of the polygons, $V_0$ the sum of the overlapping 
area between them,  and $V_b$ the area of the rectangular box. 
The initial volume fraction is $\Phi_0=84.49\pm0.05\%$.

Initially, the sample is isotropically  compressed until the pressure 
$p_0$ is reached. Then, the sample is subjected to vertical load-unload 
cycles  as 
$\sigma_1 = p_0 + \frac{\sqrt{2}}{2} \Delta\sigma[1+\cos(\pi t/t_0)]$ 
whereas $\sigma_2 = p_0 $ is kept constant.
Part (a) of Fig. \ref{mosaico} shows the relation between the shear
stress $q=(\sigma_1+\sigma_2)/2$ and the shear strain $\gamma$ 
in  the case of a loading amplitude $ \Delta\sigma=0.6p_0$. 
This relation consists of open  hysteresis loops which narrows as consecutive 
load-unload cycles are applied.  This hysteresis produces an
accumulation of strain with the number of cycles which is represented by  
$\gamma_N$ in the part (b) of Fig \ref{mosaico}. 
We observe that the strain response consists of  short time regimes, with 
rapid accumulation of plastic strain, and long time {\it ratcheting} regimes, 
with a constant accumulation rate of plastic strain of around 
$2.4\times 10^{-6}$ per cycle. 

Part (c) of Fig. \ref{mosaico} shows the relation between 
the shear stress and the volume fraction. This consists of asymmetric 
compaction-dilation cycles which makes the sample to compact during
the cyclic loading. This compaction is shown in part (d) of the 
Fig. \ref{mosaico}. We observe a slow variation  of the volume fraction 
during the {\it ratcheting} regime, and a  rapid  compaction during the 
the transition between two ratcheting regimes.   
The evolution of the volume ratio seems to be  
rather sensitive to the initial random structure of the polygons. 
Even so we found that after  $8 \times 10^3$  cycles the volume fraction 
still slowly increasing in all the samples, without any evidence of a 
saturation level.

\begin{figure}[h]
\begin{center}
 \epsfig{file=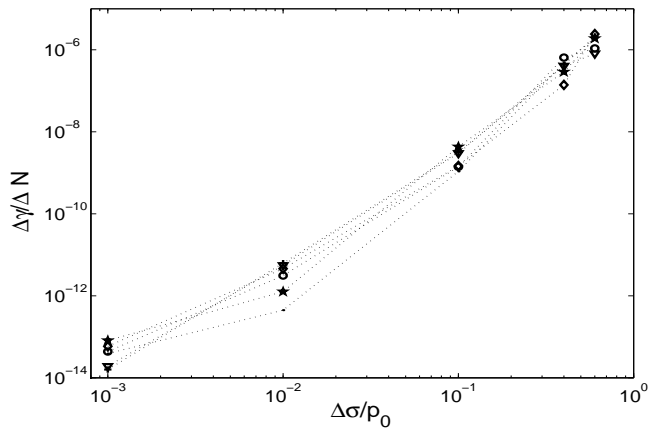,height=0.65\linewidth,width=1.0\linewidth,angle=0,clip=1}
 \end{center}
\caption{Plastic deformation per cycle 
for different loading amplitudes . The calculations 
are performed on six different samples.}
\label{power}
\end{figure}

\begin{figure}[h]
\begin{center}
 \epsfig{file=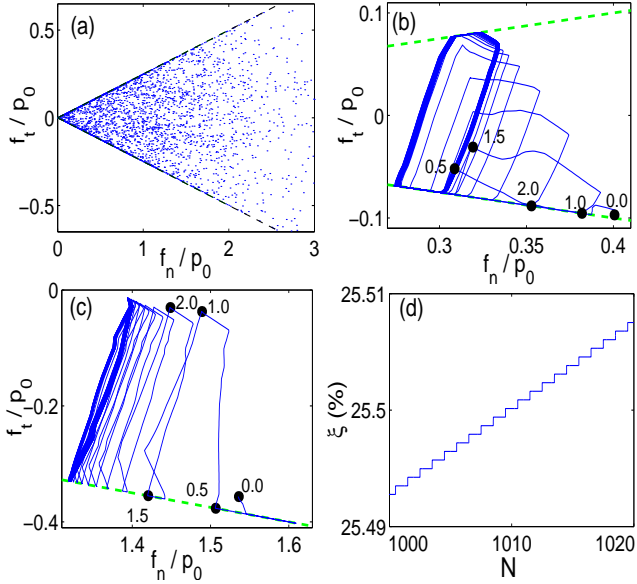,height=0.9\linewidth,width=\linewidth,angle=0,clip=1}
\end{center}
\caption{ (a) Each point represents the normal and tangential force at one 
contact. The dotted line represents the sliding
condition $|f_t|=\mu f_n$. (b) and (c) Trajectories of the contact force
of two selected sliding contacts. The dots denote 
the times $t=0,0.5t_0,...,2t_0$ in unit of the period $t_0$. 
(d) Plastic deformation $\xi$ at the contact shown in (c).}

\label{force_loops}
\end{figure}

This extremely slow dynamics in the evolution of the granular packing
shows an astonishing analogy with the behavior of glassy systems  
\cite{liu98,nicolas01}. Even more surprising is that no elastic
regime is detected by decreasing the amplitude of the loading 
cycles. Fig. \ref{power} shows the accumulation rate of strain
$\Delta\gamma/\Delta N$  for different loading  amplitudes 
$\Delta \sigma$. A constant accumulation of strain is observed 
during the cyclic loading, even when the amplitude is as slow as 
$10^{-3}$ times the applied pressure.

The existence of these ratcheting regimes proves to be a consequence of
the particular disorder of the distribution of contact forces. It is well know 
that  the stress  applied on the boundary is transmitted through an 
heterogeneous  network of contact forces \cite{jaeger96}.  
Both tangential and normal force distributions 
turn out to be very broad.  The part (a) of Fig. \ref{force_loops} shows 
the distribution of  contact forces in a polygonal packing that has been 
isotropically  compressed.  The components of the force are limited by the 
condition $|f_t| \le \mu f_n$, where $\mu$ is the friction coefficient. 
Most of the contacts satisfy the elastic condition $|f_t| < \mu f_n$,
but due to the heterogenieties,  some contacts are able to reach the 
sliding condition $|f_t| = \mu f_n$ during the compression. 
Under arbitrarily small load stress cycles these contacts
slide, given irreversible motion between the grains. Those sliding 
contacts are the most relevant micromechanic  rearrangements during the 
cyclic loading. Opening and closure of contacts are quite rare events, 
and the coordination number of the packing keeps approximately it initial 
value $4.43\pm0.08$  in all the simulations.

After some loading cycles a quasi-periodic irreversible motion is observed 
at these sliding contacts.  Two different sliding modes are represented in 
Fig. \ref{force_loops}: (b) a contact sliding forward in the load phase
and backward in the the unload one and (c) a contact sliding in the load phase 
and sticking in the unload one. A measure of the  plastic deformation of the 
sliding contact is given by $\xi = (\Delta x^c_t-\Delta x^e_t)/\ell$,
where $\Delta x^c_t$ and $\Delta x^e_t$ are the total and the elastic part of 
tangential displacement at the contact, the last one being given by 
Eq. \ref{friction}. The part (d) of Fig. \ref{force_loops} shows  the 
plastic deformation $\xi$ of the sliding contact shown in (c). Due to the 
load-unload asymmetry of the contact force loop, a net accumulation 
of plastic deformation is observed in each cycle. This is given by a  
slip-stick mechanism which resembles mechanical ratchets.

\begin{figure}[h]
\begin{center}
 \epsfig{file=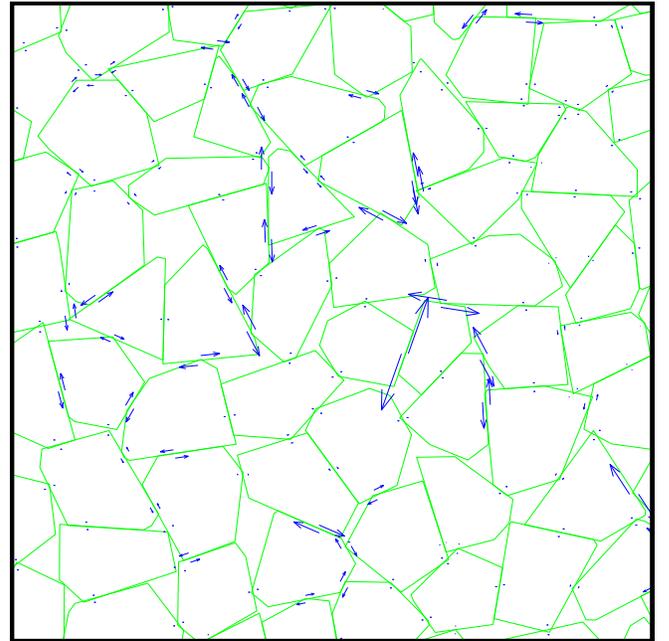,width=\linewidth,height=\linewidth,clip=1}
\caption{
The arrows represents the field  $\vec{u}$ of the plastic 
deformations accumulated at the contacts during one cycle: 
$\vec{u}=500(\xi_{N+1}-\xi_N)$, where $\xi_N$ is the plastic displacement
after $N$ cycles.}
\label{zoom}
\end{center}
\end{figure}

It is interesting to see the spatial correlation of these ratchets.
Fig. \ref{zoom} shows a snapshot of the field of plastic displacement 
per cycle at the contacts inside of the specimen. 
We see that correlated displacements coexist with a strongly inhomogeneous 
distribution of amplitudes.  Localized slip zones appear periodically  
during each ratcheting regime.  Some slip zones are  destroyed and new 
ones are created during the  transition between two ratcheting regimes. 
Moreover,  we notice that these  ratchets are found as well  at the 
boundaries as in bulk material, without  the layering 
effects  observed  in vibrated  granular materials \cite{jaeger96}.

\newpage

\begin{figure}[h]
\begin{center}
 \epsfig{file=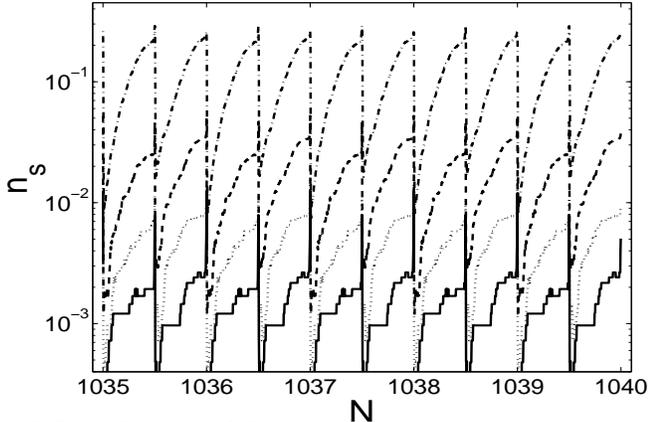,width=\linewidth,height=0.65\linewidth,clip=1}
\caption{Fraction of sliding contacts $n_s$ in the long time behavior for
different values of $\Delta \sigma/p_0$: 0.6 (dash-dotted line), 0.1 (dashed
line), 0.01 (dotted line) and 0.001 (solid line)}
\label{nc}
\end{center}
\end{figure}

Let us next discuss the correlation between the dynamics of the sliding 
contacts and the evolution of the stiffness of the material. The last one
is given by the slope of the stress strain curve in part (a) of the Fig. 
\ref{mosaico}. The evolution of the fraction  $n_s = N_s/N_c$ of sliding 
contacts with the number of loading cycles is shown in Fig. \ref{nc}.
Here $N_s$ is the number of sliding contacts and $N_c$ is the total 
number of contacts. During each loading phase, the number of sliding 
contacts increases,  giving rise to a continuous decrease of the stiffness 
as shown in part (a) of Fig. \ref{mosaico}.  An abrupt reduction in the number 
of sliding  contacts is  observed at the transition   from load to unload, 
producing a  discontinuity  in the stiffness and hence, 
a plastic deformation.
We can see also that some contacts reach almost periodically the sliding
state even for extremely small loading cycles. The ratchet-like behavior of 
these contacts produces a net displacement of the hysteretic stress-strain 
loop in each cycle, ruling out an elastic regime.

The fact that there is no evidence of an elastic regime in the
cyclic loading response suggests that the hysteresis of the 
granular materials can not be described by using the
classical theory of elasto-plasticity. Actually, the deficiencies 
of this theory have been addressed in the experimental investigation 
of soil deformation \cite{gudehus84}. Ten years ago, the 
{\it hypoplastic model} was formulated  in order to mend these 
deficiencies. This  continuum theory has been supported on the experimental 
evidence that any load involves plastic deformation \cite{kolymbas91}.  

The hypoplastic approach of granular materials requires 
to introduce new ingredients in the  current micro-mechanical description. 
The traditional fabric tensor, measuring the distribution of the orientation 
of the contacts,  can not fulfill this description, because it does not make a
distinction between  elastic and sliding contacts \cite{rothenburg88}. 
New structure tensors taking into account the statistics of the 
sliding contacts, must be introduced in order to give a 
micro-mechanical basis to the granular ratcheting. The identification
of these internal variables and the determination of their evolution equations 
and their connection with the macroscopic variables would be an important key 
to develop an appropriate continuous description of granular soils.

To conclude, we have performed a grain scale investigation of the 
cyclic loading response of a polygonal packing. 
We have shown the existence of long time 
regimes with a constant accumulation of plastic deformation per cycle, due to 
ratcheting motion at the sliding contacts.  This phenomenon may have deep 
implications in the study of the permanent deformation of soils subjected to 
cyclic  loading. More precisely, it may be necessary to introduce internal 
variables in the  constitutive  relations,  connecting  the dynamics  of 
the sliding  contacts with the  evolution of the continuous variables 
during cyclic loading. 

At this time, a comparison of the dynamic simulations with realistic 
situations  is limited by the computer time of the simulations.  
Using a computer of  $2.4GHz$ we are able  to simulate only  
$3\times10^3$ cycles per week.  Such amount of cycles corresponds only to 
some hours in realistic  cases such as railroads and 
highways \cite{lekarp00}. However, further investigations of the 
role of the stiffness, the initial stresses, and the inertial effects on 
this ratcheting behavior are currently feasible.


We thank G. Gudehus, P. Cundall, D. Potyondy, A. Schuenemann, J. Gallas and 
S. McNamara for helpful  discussions and acknowledge the support of the
DFG project {\it Modellierung koh\"asiver Reibungsmaterialen\/} and the
European DIGA project HPRN-CT-2002-00220.

\end{multicols}

\end{document}